   \documentclass[preprint,review,12pt]{elsarticle}

\usepackage{graphicx}
\graphicspath{{./}{../figure/}}
\DeclareGraphicsExtensions{.eps}

\usepackage{amsmath,amssymb}

\usepackage{color}
\usepackage{bm}
\newcommand{\kp}{\mbox{$\bm{k} \! \cdot \! \bm{p}$}}

\journal{Computer Physics Communications}

\begin{document}

\begin{frontmatter}



\title{Full-band electronic structure calculation of semiconductor nanostructures: a reduced-order approach}


\author[polito,cnr,bu]{Francesco Bertazzi}
\ead{francesco.bertazzi@polito.it}
\author[polito]{Xiangyu Zhou}
\author[polito,cnr]{Michele Goano}
\author[bu]{Enrico Bellotti}
\author[polito]{Giovanni Ghione}

\address[polito]{Dipartimento di Elettronica e Telecomunicazioni, Politecnico di Torino, corso Duca degli Abruzzi 24, 10129 Torino, Italy}
\address[cnr]{IEIIT-CNR, Politecnico di Torino, corso Duca degli Abruzzi 24, 10129 Torino, Italy}
\address[bu]{ECE Department, Boston University, 8 Saint Mary's Street, 02215 Boston, MA}

\begin{abstract}
We propose an efficient reduced-order technique for electronic structure calculations of semiconductor nanostructures, suited for inclusion in full-band quantum transport simulators.
The model is based on the linear combination of bulk bands obtained by the empirical pseudopotential method, combined with the use of problem-matched basis functions numerically generated from the singular value decomposition.
The efficiency and accuracy of the proposed approach are demonstrated in the case of the dispersion relation of hole subbands in an unstrained GaN layer.
\end{abstract}


\end{frontmatter}



\section{Introduction}

A rigorous atomistic description of the electronic structure in confined systems is of crucial importance to understand quantum transport phenomena in technologically relevant nanostructures \cite{2003DiCarlo_SST}.
Traditionally, theoretical studies of nanostructures have relied on the multiband \kp\ method \cite{2009LewYanVoon} in the framework of envelope-function approximation (EFA) \cite{1990Smith_RMP}.
In this approach the electronic states of the nanostructure are expanded in zone-center Bloch wavefunctions of the underlying bulk crystal, and the expansion coefficients (envelope functions) are assumed to be slowly varying spatial functions.
Despite the numerous approximations involved, envelope function approaches based on \kp\ models have been widely adopted, mainly due to a fair compromise between simplicity and reliability.
However, there is evidence that a full-zone description is critical for an accurate quantitative modeling of nanostructures.
As an example, quantum-mechanical mixing between the zone center $\Gamma$ and the zone edge $X$ states ($\Gamma$--$X$ coupling) due to the crossover from direct to indirect band gap in some cubic III-V systems has a relevant effect on electrical and optical properties \cite{1997Wang_PRL}.
In general, full-zone electronic structure models are attractive because they allow for an atomistic description of the band structure and a unified treatment of bound and unbound states.
The latter feature is important in the description of scattering mechanisms between continuum and bound states \cite{1986Brum_PRB, 1996Abou-Khalil_PTL, 1997Register_APL}, which play a crucial role e.g.\ in the dynamic properties of semiconductor lasers and in the quantum efficiency of light-emitting diodes (LEDs).
As a topical example, the efficiency droop observed in GaN-based LEDs \cite{2010Piprek_PSSA} has been attributed by some researchers to Auger recombination processes which, due to the large energy gap, promote carriers to high energy states in different bands above the barrier, thus contributing to leakage \cite{1990Chik_JAP, 2012Bertazzi_APL}.
A correct microscopic description of such interband processes would require a full-zone approach.
An atomistic description of the band structure is also important for the correct inclusion of strain and of the strong piezoelectric charges which are predicted in GaN-based heterostructures \cite{2007Bernardini}.
The conventional EFA approach is oblivious of the atomic details of heterointerfaces and implies a slow variation of the envelope function.
While interface-related effects lying outside the scope of conventional EFA can be accounted for, in principle, by exact envelope function theories such as those developed by Burt \cite{1992Burt_JPCM} and Foreman \cite{1997Foreman_PRB}, the description of realistic strain configurations in partially relaxed structures and the treatment of defects call for an atomistic approach.

An alternative to EFA, along the lines of the linear combination of atomic orbitals (LCAO) method \cite{1954Slater_PR,1999Munoz_CPC,2011Schulz_PSSB}, is obtained by the linear combination of bulk bands (LCBB) \cite{1999WangLW_PRB.1, 2005Esseni_PRB}, which avoids the decomposition of the wavefunction into envelope functions by expanding the states of the quantum structure in terms of the full-zone Bloch eigenstates of the constituent bulk crystals.
LCBB guarantees that the physical symmetry of the system is preserved, and allows for an atomistic description of surfaces, interfaces, and strain \cite{1999WangLW_PRB.1}.
Compared with exact diagonalization techniques, LCBB enables to select the physically important bands and wavevector points \cite{1999WangLW_PRB.1}.
As a result, the number of basis functions can be reduced significantly compared to the plane-wave basis.
Although LCBB has been applied to large scale electronic structure calculations \cite{1997Wang_PRL}, the method is still too computationally intensive to be included in carrier transport device simulation codes.
This applies in particular to nitride-based nanostructures, where charge rearrangement induced by the presence of externally applied or internally induced fields has to be considered for a realistic device description \cite{2001Chirico_PRB}.
Ideally, one should try to combine a complete quantum-mechanical description with a full-band approach, avoiding the computationally prohibitive load of atomistic methods and the inherent limitations of the EFA.

\section{From LCBB to LCBB-SVD}

With a view of the above remarks, we present a simple acceleration strategy, LCBB-SVD, based on the singular value decomposition (SVD).
This strategy was inspired by a numerical technique proposed by some of the authors to generate reduced sets of problem-matched basis functions in guided-wave finite-element analysis \cite{2002Bertazzi_MTT, 2003Bertazzi_MTT} and already demonstrated for bulk electronic structure calculations \cite{2010Penna_IWCE}.
Although this reduced-order technique can be applied to any type of nanostructure (2D, 1D, or 0D), here we restrict our attention to 2D systems with a confining potential $U(\bm{r},z)=U(z)$ which is constant in the $\bm{r}$ plane normal to the confining direction $z$.
Before presenting the details of the reduced-order model, we briefly summarize LCBB.
Following the notation in \cite{2005Esseni_PRB}, we write wavevectors and reciprocal lattice vectors as $\bm{K}=(\bm{k},k_z)$, $\bm{G}=(\bm{g},g_z)$, respectively.
We assume that the nanostructure is composed of a single material%
\footnote{This approximation is well justified for semiconductor-insulator heterojunctions where the band discontinuity is very large, or for the opposite extreme of weakly confining quantum wells (QWs).
For vertical transport across III-nitride multiple QWs, where this simplification could be questionable, the proposed method can be extended to take into account the full details of realistic heterojunctions.}
and we expand the nanostructure eigenfunction $\psi$ as a linear combination of bulk states $\Phi_{n'\bm{k}'k_z'}$ of the material considered
\begin{equation}
\psi = \sum_{n'\bm{k}'k_z'}{A_{n'\bm{k}'k_z'}\Phi_{n'\bm{k}'k_z'}}
\label{psi}
\end{equation}
where $\Phi_{n'\bm{k}'k_z'} = u_{n'\bm{k}'k_z'}(\bm{r},z)e^{j\bm{k}'\cdot \bm{r}}e^{jk_z' z}$
and $u_{n'\bm{k}'k_z'}$ is the periodic component, which can be expanded in the plane wave basis set with coefficients $B_{n'\bm{k}'k_z}(\bm{g},g_z)$
\begin{equation}
u_{n'\bm{k}'k_z'}(\bm{r},z) = \frac{1}{\sqrt{V}} \sum_{\bm{g},g_z} B_{n'\bm{k}'k_z'}(\bm{g},g_z) e^{j\bm{g}\cdot \bm{r}} e^{jk_z' z}.
\end{equation}
The unknown wavefunction $\psi$ must satisfy the Schr\"odinger equation $[H+U(\bm{r},z)]\psi = \epsilon \psi$, where $H$ is the Hamiltonian corresponding to the kinetic energy operator and the periodic crystalline potential.
By projecting the Schr\"odinger equation on the generic state $\Phi_{n\bm{k}k_z}$, a separate eigenvalue equation for each wavevector $\bm{k}$ in the unconstrained plane can be obtained with an appropriate choice of the expansion volume in the reciprocal lattice space
\begin{equation}
\begin{split}
E_{FB}^{(n)} (\bm{k},k_z) A_{n\bm{k}k_z} + \nonumber \\
\frac{2\pi}{L}\sum_{n',k_z'} \sum_{\bm{G}_z} U_T(k_z'-k_z+G_z)S_{\bm{k}k_z\bm{k} k_z'}^{(n,n')}(\bm{G}_z)A_{n'\bm{k}k_z'} =  \nonumber \\
\epsilon(\bm{k})A_{n\bm{k}k_z}
\end{split}
\label{eig}
\end{equation}
where $L={2\pi}/{\Delta k_z}$ is the length of the nanostructure in the $z$ direction, $\Delta k_z$ is the spacing between $k_z$ values, $U_T(q_z)$ denotes the 1D Fourier transform of the confining potential $U(z)$, $\bm{G}_{z}$ is a reciprocal lattice vector along $k_z$, and $S_{\bm{K}\bm{K}'}^{(n,n')}(\bm{G}_z)=\langle u_{n,\bm{K}+\bm{G}_z} | u_{n',\bm{K}'}\rangle $ are overlap integrals between periodic components.
The wavevector $(\bm{k},k_z)$ belongs to the first 2D Brillouin zone \cite{2005Esseni_PRB}, thus $|k_z|<G_{zm}/2$, where $G_{zm}$ is the magnitude of the smallest reciprocal lattice vector $\bm{G}_z$.
The empirical pseudopotential method (EPM) \cite{1988Cohen} is used to determine both the full-band dispersion $E_{FB}^{(n)} (\bm{k},k_z)$ and the Fourier components $B_{n\bm{k}k_z}(\bm{g},g_z)$ of the periodic functions of the underlying 3D crystal.
The expansion coefficients $B_{n\bm{k}k_z}(\bm{g},g_z)$ are then used to compute the overlap integrals $S_{\bm{k}k_z\bm{k} k_z'}^{(n,n')}(\bm{G}_z)$.
The sum over $\bm{G}_z$ in (\ref{eig}) can be safely truncated to include only the terms $\bm{G}_z=(\bm{0},\pm G_{zm})$ in addition to $\bm{G}_z=(\bm{0},0)$ \cite{2005Esseni_PRB}.

LCBB leads to an eigenvalue problem of rank $n_b n_{k_z}$, where $n_{k_z}$ is the number of wavevectors along $k_z$ and $n_b$ is the number of bulk bands included in the calculation.
The number of points $n_{k_z}$ in (\ref{psi}) needed to achieve a given accuracy is structure-dependent \cite{2001Chirico_PRB}.
Moreover, the presence of a confining potential implies a lower bound on the number of points for the Fourier representation of the potential itself.
In general, a few hundred points are necessary for an accurate description of the energy dispersion of typical 2D systems.
In practical cases, the computation time is dominated by the calculation of the matrix itself (which requires the calculation of the Bloch waves $\Phi_{n\bm{k}k_z}$) rather than by its diagonalization.
We will show that the information necessary to build the eigenvalue matrix (\ref{eig}) can be efficiently extracted from a few numerically generated problem-matched basis functions.
To this end, it is convenient to write the Hamiltonian $H$ in the form used in complex band structure calculations \cite{1990Smith_RMP}, which explicitly displays the $k_z$ dependence of $H$ (non-local terms and spin-orbit corrections are not included for conceptual simplicity)
\begin{equation}
H_{\bm{G},\bm{G}'}(\bm{k},k_z) = H^2_{\bm{G},\bm{G}'}k^2_z + H^1_{\bm{G},\bm{G}'}k^2_z + H^0_{\bm{G},\bm{G}'}
\label{Smith}
\end{equation}
with
\begin{align}
H^2_{\bm{G},\bm{G}'} &= \frac{\hbar^2}{2m} \delta_{\bm{G},\bm{G}'} \\
H^1_{\bm{G},\bm{G}'} &= \frac{\hbar^2}{m} g_z \delta_{\bm{G},\bm{G}'} \\
H^0_{\bm{G},\bm{G}'} &= \frac{\hbar^2}{2m}(k^2+2\bm{k}\cdot \bm{g} +g^2) \delta_{\bm{G},\bm{G}'} + V(\bm{G}-\bm{G}')
\end{align}
Rather than repeatedly solving the bulk problem for each $(\bm{k},k_z)$ point, we diagonalize $H$ in a few selected $n_p$ points (the \emph{expansion points}) evenly spaced along $k_z$ between $0$ and $G_{zm}/2$.
The eigenvectors computed at the expansion points are then arranged columnwise in a matrix $X$ with dimensions $n_G \times n_b n_p$, where $n_G$ is the number of plane waves used in the EPM calculations.
By applying the economy-size SVD \cite{96Golub} $X = U \Sigma V^{\dagger}$, we obtain a $n_G \times n_b n_p$ unitary matrix of left singular vectors $U$, a $n_b n_p \times n_b n_p$ diagonal matrix $\Sigma$ with positive elements (the singular values), and a $n_b n_p \times n_b n_p$ unitary matrix of right singular vectors $V$.
The significance of each singular vector in the description of the bands considered is measured by the amplitude of the corresponding singular value \cite{96Golub}.
Since singular values typically range over several orders of magnitude, just a few of them may be needed to obtain an accurate description of the band structure.
Having selected a suitable lower bound for the singular values, the bulk problem can be efficiently solved for arbitrary points in the range $0<k_z<G_{zm}/2$ by diagonalizing the reduced-order Hamiltonian of rank $\tilde{n}_{G}$
\begin{align}
\tilde{H}_{\bm{G},\bm{G}'}(\bm{k},k_z) = \tilde{H}^2_{\bm{G},\bm{G}'}k^2_z + \tilde{H}^1_{\bm{G},\bm{G}'}k_z + \tilde{H}^0_{\bm{G},\bm{G}'}
\label{reduced_Smith}
\end{align}
where $\tilde{H}^{\alpha}_{\bm{G},\bm{G}'} = \hat{U}^{\dagger} H^{\alpha}_{\bm{G},\bm{G}'} \hat{U}$ with $\alpha=0,1,2$ and the columns of $\hat{U}$ are the $\tilde{n}_{G}$ columns of $U$ corresponding to singular values larger than the lower bound.
The eigenvectors used to build matrix $X$ should be carefully selected in order to prevent spurious bands in the desired energy range of the dispersion relation.
As a general rule, ghost solutions may appear if bands lying close to the energy range of interest are not included in $X$.
Having computed the Bloch states in the interval $0<k_z<G_{zm}/2$ with the reduced-order model, the additional information needed in the secular equation (\ref{eig}) can be obtained by symmetry considerations.
The wavefunction coefficients associated to $(\bm{k},k_z)$ and $(\bm{k},-k_z)$ are related (to within a phase) by \cite{1990Smith_RMP}
\begin{equation}
B_{n\overline{\bm{K}}}(\bm{G}) = B^*_{n,\bm{K}}(\overline{\bm{G}})
\label{inv}
\end{equation}
with $\overline{\bm{K}} = (\bm{k},-k_z)$ and $\overline{\bm{G}} = (\bm{g},-g_z)$.
A phase factor may apply depending of the specific choice of the primitive vectors \cite[App.~A]{04Hjelm_PhD}, \cite{2000Brennan_TED}.
In the reduced-order representation, Eq.~(\ref{inv}) can be written as $\tilde{B}_{n,\tilde{\bm{K}}} = \tilde{T}_{\bm{G},\tilde{\bm{G}}} \tilde{B}^*_{n,\bm{K}}$, where $\tilde{T}_{\bm{G},\tilde{\bm{G}}}=\hat{U}^{\dagger} T_{\bm{G},\tilde{\bm{G}}} \hat{U}^*$ and $T_{\bm{G},\tilde{\bm{G}}}$ is the matrix that incorporates the swap sequence of the $\bm{G}$ vectors with the appropriate phase factors.
Out-of-zone states can be constructed using the periodicity condition (strictly valid if the basis set is not truncated)
\begin{equation}
B_{n,\bm{K}+\bm{G}_z}(\bm{G}) = B_{n,\bm{K}}(\bm{G}+\bm{G}_z)
\end{equation}
leading to $\tilde{B}_{n,\bm{K}+\bm{G}_z} = \tilde{T}_{\bm{G},\bm{G}'+\bm{G}_z} \tilde{B}_{n,\bm{K}}$, with $\tilde{T}_{\bm{G},\bm{G}'+\bm{G}_z}=\hat{U}^{\dagger} T_{\bm{G},\bm{G}'+\bm{G}_z} \hat{U}$ and $T_{\bm{G},\bm{G}'+\bm{G}_z}=\delta(\bm{G},\bm{G}'+\bm{G}_z)$.
Once all the necessary bulk eigensolutions have been computed, the overlap integrals $S_{\bm{k}k_z\bm{k} k_z'}^{(n,n')}(\bm{G}_z)$ can be efficiently computed by evaluating scalar products in the reduced-order subspace spanned by $\hat{U}$.

\begin{figure}
\centerline{\includegraphics[width=1.0\columnwidth]{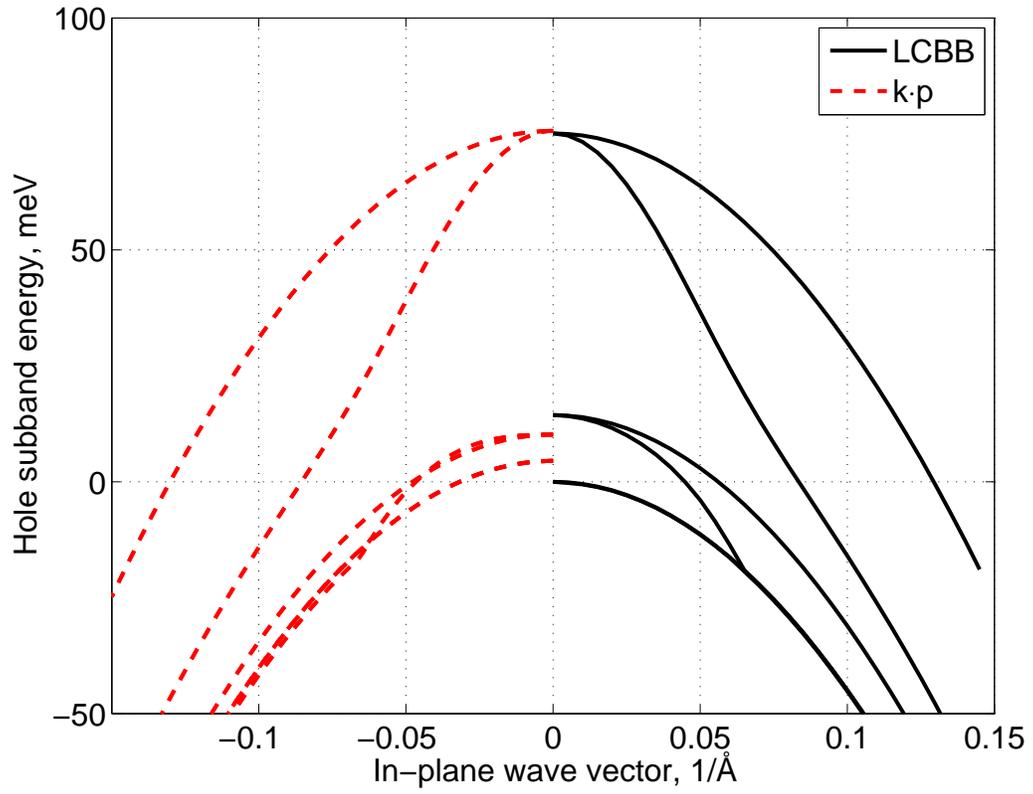}}
\caption{Valence subband structure of an unstrained 25\,\AA\ thick GaN layer with a confining potential of 0.1\,eV along [001], computed with LCBB (solid lines) and \kp\ EFA (dashed lines).}
\label{GaN001}
\end{figure}

\begin{figure}
\centerline{\includegraphics[width=1.0\columnwidth]{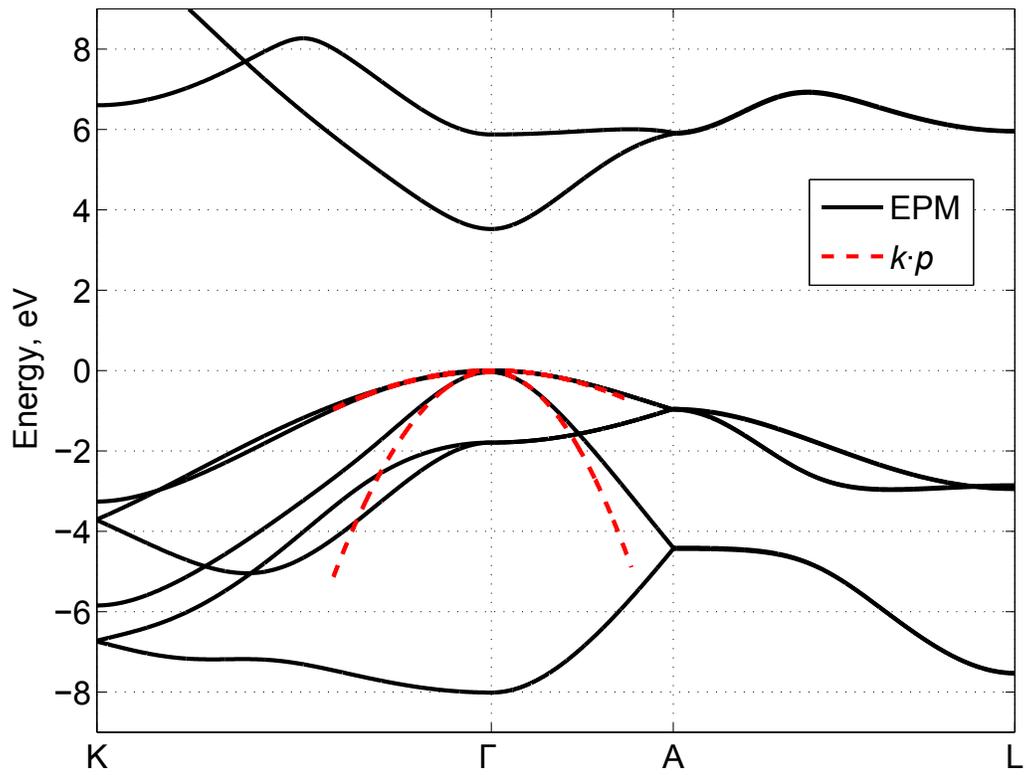}}
\caption{Electronic structure of wurtzite GaN computed with EPM (solid lines) and its $6\times 6$ \kp\ fit (dotted lines).}
\label{EPMvskp}
\end{figure}

\section{Application example}

Active regions of blue- and green-emitting optoelectronic devices usually consist of multiple III-nitride QWs.
As a numerical example, we calculated the in-plane hole subband structure of an unstrained 25\,\AA\ thick GaN layer with a confining potential of 0.1\,eV along [001].
The LCBB-SVD model was obtained by computing the upper six valence bands at $n_p=5$ expansion points, evenly spaced between 0 and $G_{zm}/2$ along $k_z$.
The EPM parameters were taken from \cite{2000Goano_JAP.1, 2007Bellotti_JAP}, with an energy cutoff corresponding to $n_G=197$ plane waves.
The application of SVD decomposition yields $n_b n_p = 30$ singular vectors.
The corresponding singular values span five orders of magnitude, which confirms that the sampling rate in momentum space is sufficient to represent the selected valence bands.
In order to eliminate the redundancy from the basis functions set, the $\tilde{n}_{G}=18$ singular vectors corresponding to singular values having magnitude larger than 1/100 of the dominant one were included in $\hat{U}$.
The reduced-order model was solved in $n_{k_z}=201$ points along the $k_z$ direction between $-G_{zm}/2$ and $G_{zm}/2$.
The selected problem-matched basis functions alone are sufficient to reproduce the LCBB dispersion diagram of the structure with excellent accuracy, and no difference can be appreciated in Fig.\,\ref{GaN001} (solid lines) between LCBB and LCBB-SVD results.
A comparison between Fortran implementations of the standard LCBB and the reduced-order technique confirms that the latter is about 10 times faster than the former.

It is interesting to compare the LCBB results with conventional multiband \kp\ EFA models.
Common \kp\ implementations for GaN are based on the symmetrized version of the wurtzite Hamiltonian \cite{1996Chuang_PRB, 1997Chuang_SST}, which is numerically unstable in structures having significant discontinuities of the material parameters at interfaces \cite{1999Mireles_PRB, 2000Mireles_PRB}.
Another important issue in EFA is the correct ordering of the differential operators.
Reliable and spurious-solution-free subband structures are calculated here with a finite element discretization in real space that includes Burt-Foreman operator ordering \cite{1992Burt_JPCM, 1993Foreman_PRB, 1997Foreman_PRB}, ensuring the ellipticity of the equations in the framework of standard EFA \cite{2008Veprek_JCE, 2009Veprek_OQE}.
Fig.\,\ref{EPMvskp} compares the electronic structure of unstrained bulk GaN computed with EPM (solid lines) and a $6\times 6$ \kp\ fitting near $\Gamma$ (dashed lines) obtained by least-squares optimization.
From Fig.\,\ref{GaN001} (dashed lines), it can be seen that $6 \times 6$ \kp\ models are able to approximate the bound states of the nanostructure in the limits of EFA.
However, simplified approaches, where bound levels are treated as 2D states within EFA while the continuum portion of the spectrum is described through bulk states, are intrinsically unable to provide a microscopic description of capture (continuum-to-bound) and escape (bound-to-continuum) processes, since initial and final states are not treated on equal footing \cite{Rossi}.
Investigations of capture processes in separate confinement heterostructures (SCH) based on \kp\ EFA and Fermi golden rule have also led to unphysical results due to the finite coherence length of the carriers \cite{1999Mosko_SST, 1999Zakhleniuk_PSSA, 1999Levetas_PRB}.
Moreover, a zone-center description of the electronic structure is not suitable to describe scattering processes that involve states far from $\Gamma$, a notable example being Auger recombination in wide band gap semiconductors \cite{2010Bertazzi_APL, 2012Bertazzi_APL}.
Although this problem can be alleviated by full-zone \kp\ approaches \cite{2004Beresford_JAP}, they usually require multiple expansion points to cover the entire Brillouin zone with a computational cost comparable to EPM and possible added complexity related to interpolation issues \cite{2007Persson_CPC, 2010Marnetto_JAP}.
The present LCCB-SVD model allows for a unified full-zone treatment of extended (bulk-like) and localized (QW-like) states with a small overhead with respect to the EPM problem for bulk semiconductors.
LCCB-SVD is therefore suited for the evaluation of scattering rates between extended and localized states, a crucial ingredient to investigate vertical carrier transport across heterostructures.

\section*{Acknowledgments}

The authors would like to thank Dr.~Michele Penna for useful discussions.
This work was supported in part by the U.S.\ Army Research Laboratory through the Collaborative Research Alliance (CRA) for MultiScale multidisciplinary Modeling of Electronic materials (MSME).







\end{document}